\documentclass[english]{maciarticle}
\usepackage{amsmath,amsbsy,amscd,amssymb,graphicx, epsfig,color,xcolor,placeins}
\begin{document}

\newcommand{\SubItem}[1]{
    {\setlength\itemindent{15pt} \item[-] #1}}

\title{Delta hedging with transaction costs: dynamic multi-scale strategy using neural nets}

\author[$\dag$]{Gaston Mazzei}
\author[$\dag$]{Fernando Bellora}
\author[ ]{Juan Serur $\ddag$ \\ 
\small{\textcolor{gray}{\emph{gastonmazzei95@gmail.com} $|$ \emph{fernando.bellora@gmail.com} $|$ {\it juan.serur@nyu.edu}}}}

\affil[$\dag$]{Universidad de Buenos Aires,
Buenos Aires, Intendente Guiraldes 2160, Argentina, 
}
\affil[$\ddag$]{New York University (NYU) - Courant Institute of Mathematical Sciences.}


\maketitle

\vspace{-0.8cm}
\begin{abstract}
In most real scenarios the construction of a risk-neutral portfolio must be performed in discrete time and with transaction costs.
Two human-imposed constraints are the risk-aversion and the profit-maximization, which together define a nonlinear optimization problem with a 
model-dependent solution.
%
In this context, an optimal fixed-frequency hedging strategy can be determined \textit{a posteriori} by maximizing a sharpe-ratio-simil path-dependent reward function. 
Sampling from Heston processes, a convolutional neural network was trained to infer which period is optimal using partial information, thus leading to a dynamic hedging strategy in which the portfolio is hedged at various frequencies, each weighted by the network's probability estimate of that frequency being optimal.  
\end{abstract}

\begin{mathsubclass}
91G - 91G20
\end{mathsubclass}

{\thispagestyle{empty}}


\section{Introduction}
\label{Intro}
\medskip
%
%

A purely risk-neutral strategy with stocks and options is a utopia, but the continuous-time hypothesis in which stochastic differential equations (SDE) are generally defined is not to blame.
What prevents the most general user from implementing continuous hedging strategies derived from the Black Scholes (BS) model \cite{BS} is the existence of transaction costs.
Intuitively, the will of updating the portfolio as much as possible in an attempt to reduce the risk opposes the will of updating the portfolio as little as possible to prevent the cash loss due to transaction costs: 
quantifying this trade-off requires modelling both risk and profit. 

The mission of this work is to build a dynamic multi-scale hedging strategy in which the weights are computed in real-time with partial information by a neural network.
That means hedging different fractions of the portfolio at different frequencies, each weighted by a neural-network's probability estimate of that frequency being optimal under a modelled reward function.
For all purposes, the considered portfolio will be that of Black and Scholes \cite{BS}, shown in Eq. \ref{portfolio}.

Previous related works include \textit{Deeep Hedging} by \textit{Bühler et al.}\cite{deephedging}, a framework in which the convergence of recurrent neural nets to optimal multi-asset trading strategies under convex measures of reward has been proven. The present work replaces the recurrent architecture with a set of convolutional networks, the need for estimating the market's volatility with using as input the directly observable derivative's market price, and the direct estimation of the asset position with the estimation of the weights of its partial fourier series.

\begin{equation}
\label{portfolio}
    Portfolio \equiv \Pi = C_{(t)} - \Delta_{(t)} S_{(t)}.
\end{equation}


\section{Proposed reward function}

\medskip
%
%
%

The trade-off between risk-aversion and profit-maximization was modeled by a path-dependent reward function shown in Eq. \ref{reward-motivation}. Modelling the risk as the squared root of the variance and using the portoflio shown in Eq. \ref{portfolio} yielded Eq. \ref{reward-motivation2}, where ``$r$'' the is the risk free rate, ``$\gamma$'' the risk aversion in arbitrary units and ``$f$'' is the percentage of each trade to deduct as transaction costs. Finally ``$T$'' is the period at which the hedging is performed, parameter against which the reward function is \textit{a posteriori} maximized.


\begin{equation}
    \label{reward-motivation}
    \left \{ \frac{\Delta \text{Reward}}{\Delta Gain} > 0 \text{ ; }    \frac{\Delta \text{Reward}}{\Delta Risk} < 0 \right \} \Rightarrow    Reward = \frac{Gain}{Risk} \approx \frac{\mathbf{E}(\Pi)}{\mathbf{E}(\sigma)},
\end{equation}


\begin{equation}
    \label{reward-motivation2}
      Reward = \frac{\Pi_{final} -\sum costs}{\gamma + \sqrt{Variance}}
    = \frac{\Pi_0(e^{rTN}-f)-f\sum_{j=1}^{N}S_{(t=jT-T)}|\Delta_{(t=jT)}-\Delta_{(t=jT-T)}|}{\gamma +\sqrt{\frac{\sum_{i=0}^{N}\sum_{j=1}^{T}([C_{(iT)}-\Delta_{(iT)}S_{(iT)}]-[C_{(iT+j)}-\Delta_{(iT+j)}S_{(iT+j)}])^2}{NT}}}.
\end{equation}

\vspace*{.1cm}

\section{Generating synthetic data by sampling from Heston processes}

\medskip
%
%
%

\begin{figure}[!ht]
\begin{center}
\includegraphics[width=0.95\textwidth]{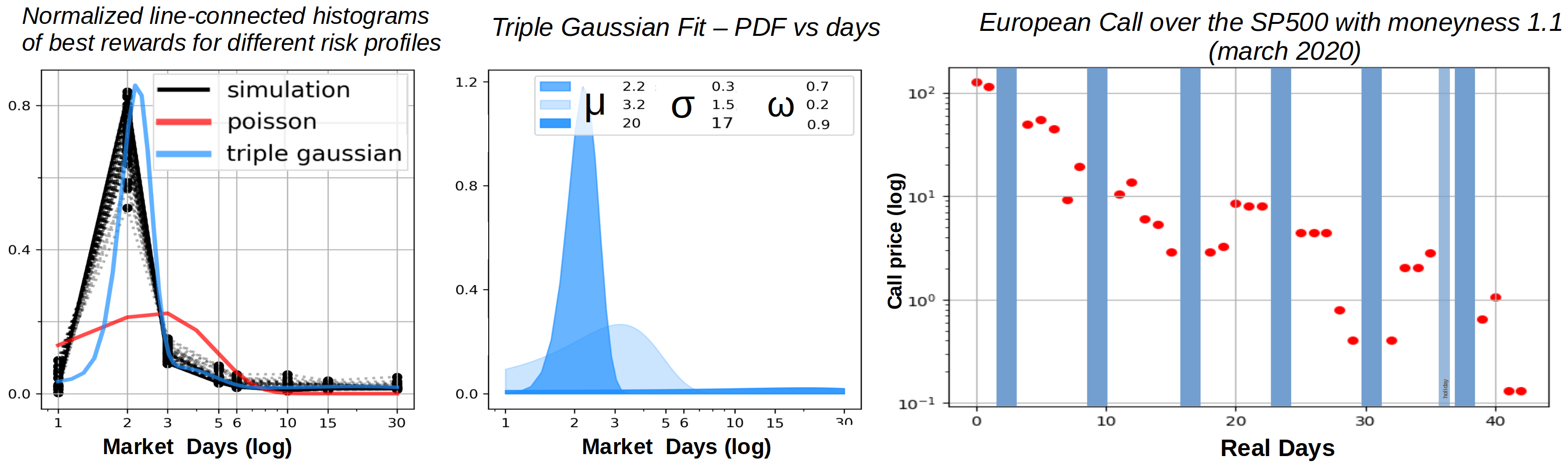}
\caption{(LEFT) for the simulated time-series, normalized histograms show the optimal trading period. Each black curve correspond to different values of risk aversion ($\gamma$) and transaction costs ($f$), both arbitrary selected over a narrow band.
(CENTER) a ternary gaussian model that can be used to (A) qualitatively show that one normally-distributed variable is not well-suited to explain the observations and (B) suggest a model with three variables for further works. (RIGHT) Real price over time for a call over the S\&P 500 Index during the Covid-19 crash (March 2020). Only 30 of the 42 days before maturity were trading days, as vertical bars show weekends and holidays.} \label{cost-info}
\end{center}
\end{figure}

Fig.\ref{cost-info} \textit{(right)} shows the time-series of a European option over the S\&P 500 index during the Covid-19 crash: while at the end real data was used to evaluate the hedging strategy, in general 1-dimensional convolutional neural networks' parameters were adjusted using synthetic data. 

First, stock prices and market volatility  over time were simulated by solving the Heston model's equations (Eq. \ref{heston-1}) using the Euler-Maruyama algorithm. 
Second, the price of the associated European Call was computed using the inverse-fourier algorithm \cite{Evaluation}. 
The first two steps allowed to compute the optimal hedging frequency for each time-series of the pair stock and European call.
To do this the transaction cost factor (``$f$'') and the risk aversion (``$\gamma$'') were fixed at ``1\%'' and ``1.5'' respectively. 
Roughly 100 instances of 1-dimensional convolutional neural networks with 3 convolutional layers and 3 dense layers (each one had about 2500 parameters) were fitted to the task of classifying, given a path up to a certain day, which hedging period is the optimal one. 
For all the simulated paths, the maturity remained fixed at 42 days and the initial moneyness at 1.1. As there are 30 trading days in 42 week days, the allowed hedging frequencies were the numbers that can perform integer division of 30: i.e. 1, 2, 3, 5, 6, 10, 15 and 30.

\begin{equation}
\label{heston-1}
    \left \{ dS_t=\mu S_tdt+\sqrt{V_t}S_tdW_t^1 \right \} \text{ ;    }
    \left \{ dV_t=a(\bar{V}-V_t)dt+\eta\sqrt{V_t}dW_t^2 \right \} \text{ ;}
    \left \{Corr(dW_t^1,dW_t^2) = \rho\right \},
\end{equation}

In Fig.\ref{cost-info} \textit{(left)} multiple histograms are plotted (black lines). Each one shows the relative frequency of each of the possible 8 periods being optimal. Each histogram was computed  over the entire dataset for different values of ``$f$'' and ``$\gamma$''. 
For comparison, Poisson \textit{(left)} and Gaussian \textit{(left and center)} fits are included in the figure.
Using maximum likelihood estimation, an alternative hypothesis to the machine-learning models was built: the probability of each period being optimal is a parameter in an input-independent random process with a multinomial distribution. The best values for the parameters are the relative frequencies for each period in the respective histogram (Eqs. \ref{mult}, \ref{periods}).

\begin{equation}
\label{mult}
        PMF(x_1=n_1,...,x_m=n_m) = \frac{\sum_{i=1}^{m}n_i}{n_1!...n_m!}p_1^{n_1}...p_m^{n_m},
\end{equation}

\begin{equation}
    \label{periods}
    \tau \text{ } \epsilon \text{ } \{1,2,3,5,6,10,15,30\} \Rightarrow m=8.
\end{equation}

\FloatBarrier

\FloatBarrier
\section{Implementing the algorithms and performing inference over real data}

The 1-dimensional convolutional neural networks' parameters were fitted with partial information, as the time-series up to certain days was used as input data. The process was repeated with two machine-learning algorithms that are not generally deemed as well-suited for the classification of time-series: a logistic regression and a random forest. The generalization error against the amount of data available is illustrated in Fig. \ref{fit}, where it is expressed in terms of the area under the ROC curve.

Finally, the dynamic multi-scaled strategy was applied: the portfolio was hedged at all the 8 possible frequencies choosing as relative weight their probabilities of being optimal. When computed over real data, the neural nets' probability estimates for each period's optimality changed over time (Fig. \ref{fit}).
Table \ref{table} shows how, by modelling transaction costs as proportional by ``1\%'', results of performing the strategy using the probability estimates of different models were derived. 

\begin{figure}[!ht]
\begin{center}
\includegraphics[width=1\textwidth]{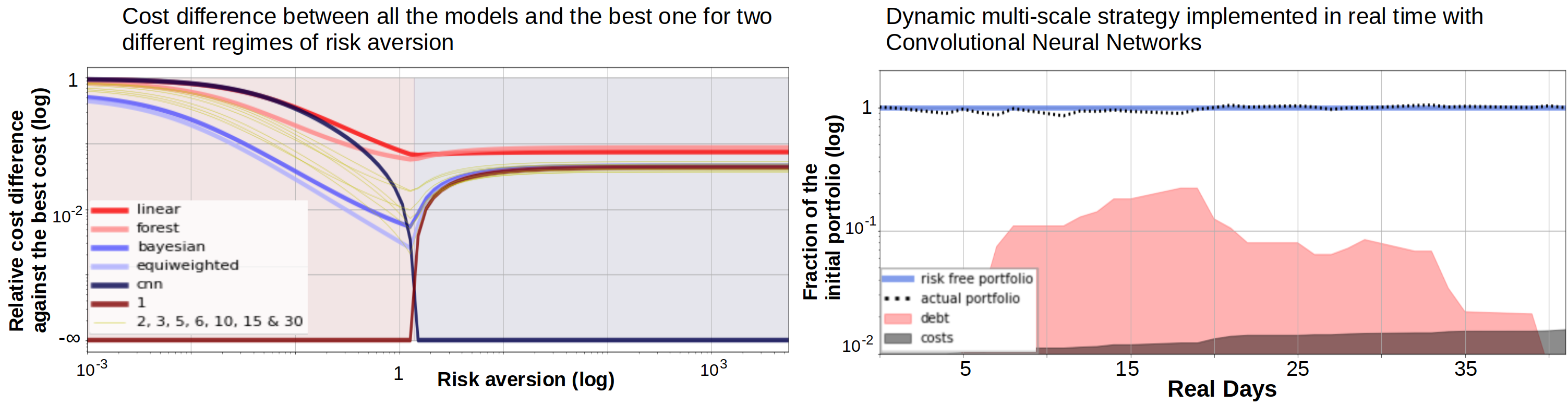}
\caption{LEFT: the reward difference between strategies and the ``best'' strategy computed against the risk-aversion. There are two regimes: when risk aversion is below $\approx1.5$ (reddish zone),  hedging every day is what maximizes the reward. For higher values (bluish zone), the dynamic multi-scale hedging performed in average between 5\% and 10\% better when using convolutional neural networks than when using the other available probability estimates. RIGHT: strategy over time  using a realistic risk-neutral rate of about 0\% per month. References: the risk-neutral portfolio (blue; solid line), debt incurred in when hedging (red; area), transaction costs (black; area), and the realized portfolio by performing the dynamic multi-scale hedging strategy using the probability-estimates of the neural networks (black; dashed).} \label{result}
\end{center}
\end{figure}

To determine which strategy performed best at maximizing the reward function is that the risk-aversion parameter (``$\gamma$'' in Eq. \ref{reward-motivation2}) was iterated over. Two regimes were distinguished: for very low levels of $\gamma$ (i.e., high levels of risk aversion) the reward-maximizing strategy is to hedge as soon as possible; in our framework this means every day. For levels of $\gamma$ higher than 1.5, the reward-maximizing strategy was the one that used the neural networks.
Results are shown in Fig. \ref{result} \textit{(left)}, along with the risk-neutral portfolio, realized portfolio, debt and transaction costs for the dynamic multi-scale strategy using the probability estimates of the neural networks \textit{(right)}.

\FloatBarrier
\begin{center}

\begin{figure}[!ht]
\includegraphics[width=1\textwidth]{classifier2.png}
\caption{
(LEFT) fitting metrics of $\approx 10$ instances of three machine-learning algorithms trained with partial information, i.e. using as input the price-curves from day zero until the horizontal value. The vertical value is the area under the ROC curves, which were computed over previously-unseen data. (RIGHT) stacked bar plot with the probabilities of each hedging period being optimal, as outputted by a 1-D convolutional neural network which was fed the time-series up to the given horizontal day. 
The predominance of short hedging periods was the model's response to an increase in volatility as the data comes from the 2020 Covid-19 Crash.
} \label{fit}
\end{figure}


\begin{table}[htp]
 \caption{results for various models over the S\&P500 Index during the Covid-19 Crash. Underlying's standard deviation in this work's units was 5.6\%, and the price variation was -10\%. Most fixed-frequency strategies replicated the mean with less variance. References are: CNN (convolutional neural network), forest (random forest classifier), linear (logistic regressor with lineal features), bayes (probabilities equal to the sample's normalized histogram), unif (every period is equally probable), \#numbers (hedging at a single constant period), Final \% (final portfolio relative to the risk neutral
 portfolio), ``Std \%'' (standard deviation of the difference between the realized portfolio and the risk neutral portfolio), ``Under \%'' (difference between the realized portfolio variation (end value vs initial value) and the variation of the underlying asset, the S\&P 500 Index). }\label{table}
\begin{tabular}
{ p{1.62cm} | p{0.7cm}|  p{0.85cm}| p{0.8cm}| p{0.8cm}|  p{0.8cm}| p{0.65cm} |  p{0.65cm} |  p{0.65cm} |  p{0.65cm}  | p{0.65cm} |  p{0.65cm}|   p{0.65cm} |  p{0.65cm} | }

  & CNN & forest & linear & bayes & unif & 1 & 2 & 3 & 5 & 6 & 10 & 15 & 30 \\
\hline
\hline
  Final (\%) & 97.4 &82.0 & 84.0 & 89.0 & 89.4 & 90.1 & 88.6 & 89.9 & 90.0 & 90.5 &90.3 & 88.3 & 87.4 \\
\hline
  Std (\%) & 12.53 & 3.76 & 10.21 & 1.03 & 0.92 & 0.00 & 1.47 & 1.97 & 1.70 & 3.24 & 3.67 & 3.88 & 2.59 \\
\hline
  Under (\%) & +4.8 & -10.6 & -8.6 & -3.6 & -3.2 & -2.5 & -4.0 & -2.7 & -2.7 & -2.1 & -2.3 & -4.3 & -5.2 \\
  
\end{tabular}
\end{table}

\end{center}

\FloatBarrier
\section{Conclusions}

The starting point was to propose a reward function that could model human-imposed constraints, and the focus was on how would a dynamic multi-scaled strategy perform compared to the optimal hedging frequency, which could only be known \textit{a posteriori}. 
The dynamic-hedging strategy was further subdivided according to which model was used for the weights-assignment: neural networks, random forests, using only one frequency, and using the bayesian posterior's expectation values among others.

Results shown in Table \ref{table} indicate that the fixed-frequency strategies failed to replicate the risk-neutral portfolio but ended replicating the underlying asset's performance with a loss of between 2\% and 4\%. They did so with half it's variance. Generic machine-learning models not suited for time-series led to the worst rewards independently of the level of risk-aversion. Finally, raw results for the strategy using neural networks show that it was both: the best risk-neutral replicator and the most risky. 
All conclusions should be tested against more data: as predicting over real data was only performed once, results lack the robustness associated with statistically significant data.
A final interesting result worth mentioning is the dynamic of the strategies as a function of the risk-aversion (Fig. \ref{result}): measured by the proposed reward function, the two optimal regimes are either hedging on a daily basis or using the dynamic multi-scale hedging strategy and computing the weights using 1-dimensional convolutional neural networks.

\end{document}